\documentclass[prb, superscriptaddress]{revtex4-2}
\usepackage{graphicx}
\usepackage{color}
\usepackage{amsmath}
\usepackage{amsfonts}
\usepackage{enumitem}
\usepackage{hyperref}
\hypersetup{
	colorlinks = true,
	citecolor = blue,
	linkcolor= blue
}
\usepackage{color}

\newcommand{\be}{\begin{equation}}
\newcommand{\ee}{\end{equation}}

\newcommand{\bea}{\begin{eqnarray}}
\newcommand{\eea}{\end{eqnarray}}

\renewcommand{\phi}{\varphi}
\renewcommand{\epsilon}{\varepsilon}

\usepackage[normalem]{ulem}

\begin{document}

\title{Nodal structure of bound-state wave functions for systems with quartic dispersion}

\author{E.V. Gorbar}
\affiliation{Department of Physics, Taras Shevchenko National University of Kyiv, 64/13, Volodymyrska Street, Kyiv 01601, Ukraine}
\affiliation{Bogolyubov Institute for Theoretical Physics, 14-b Metrolohichna Street, Kyiv 03143, Ukraine}

\author{B.E. Grinyuk}
\affiliation{Bogolyubov Institute for Theoretical Physics, 14-b Metrolohichna Street, Kyiv 03143, Ukraine}

\author{V.P. Gusynin}
\affiliation{Bogolyubov Institute for Theoretical Physics, 14-b Metrolohichna Street, Kyiv 03143, Ukraine}

\begin{abstract}

The nodal structure of bound-state wave functions for one-dimensional quantum systems with quartic energy–momentum dispersion, $E(p) \sim p^4$, and polynomial
potentials is analysed by using the semiclassical approximation and variational approach. For energies of bound states, we derive the quantization condition, obtained
by using the complex Wentzel method, where we take into account perturbative (up to the fourth order) and nonperturbative in the Planck constant corrections. The
bound-state energies and wave functions for the harmonic and quartic potentials  are compared with those found by applying the variational approach utilizing the universal
Gaussian basis. It is shown that the classical oscillation theorem, valid for systems with quadratic energy-momentum dispersion, breaks down in the classically forbidden
region where wave functions also have nodes, while it still remains valid in the classically allowed region. These results are confirmed in addition via the solutions of the
exactly solvable problem of the fourth-order Schrödinger equation with a square well potential.
\end{abstract}

\maketitle

\section{Introduction}

Quantum mechanical systems in which potential energy dominates over the kinetic energy attract a significant interest because they often exhibit strongly
correlated behavior and give rise to novel quantum phases of matter which are not well described by traditional perturbative or mean-field approaches \cite{Laad,Kuzian}.
These systems are central to many open questions in modern physics including entanglement, the simulation of strongly interacting gauge theories in condensed matter analogs,
and the realization of quantum criticality and emergent symmetries. The relevant condensed matter systems include high-temperature superconductors (cuprates), heavy fermion
systems, transition metal oxides, twisted bilayer graphene which exhibit such interaction dominant phases as correlated insulating phases, Wigner crystal, unconventional
superconductivity, and non-Fermi liquid behavior \cite{Paschen,Shayegan,Stewart,Lee}.

Quasiparticles with the dispersion relation $E(p) \sim p^{2n}$, $n\geq 2$, refer to emergent excitations in quantum many-body systems whose energy-momentum dispersion
is soft, meaning that their energy increases slowly with momentum near zero momentum. Such systems naturally realize the regime of dominance of potential energy over
the kinetic energy and for large $n$ describe weakly dispersing bands. Theoretical studies explore how even weakly dispersing bands can lead to enhanced susceptibilities, magnetism, superconductivity, or localization \cite{Flach}.

The problem of localization is related to the study of disorder and impurity induced bound states thus providing a motivation for investigating bound states in systems with quartic dispersion. An important characteristics of bound states is the number of nodes of their wave functions. The nodal structure of wave functions of bound states for systems with quartic dispersion is the main focus of the present paper.

Mathematically, finding eigenvalues and eigenfunctions of bound states for quasiparticles with soft dispersion
implies solving higher order Schrödinger-like equation in the presence of an interaction potential which is not an easy task. Recently,  two of us applied the semiclassical Wentzel-Kramers-Brillouin
(WKB) approach to determine bound state energies for quasiparticles with quartic dispersion and polynomial potentials \cite{quartic}. It should be noted that for fourth-order ordinary
differential equations with generic potential $V(x)$ only local WKB solutions in low orders in the Planck constant $\hbar$ are available in the literature (see, e.g.,
\cite{Saito1959,Zaslavskiy-book,Fedoryuk-book}), but their global properties, related to quantization, only begin
to be investigated \cite{Ito2021}. Since the WKB method is not applicable in vicinity of turning points, to match semiclassical wave functions in the classically allowed and
forbidden regions one needs to obtain the connection formulas. The latter are defined by solutions to the equation with linearised potential in vicinity of turning points which
are given by the fourth-order Airy functions \cite{Ansari2017,Dorugo_thesis} in the case of quartic dispersion.
Determining the leading asymptotics for these functions and their exponentially suppressed terms (hyperasymptotics) necessary for matching the WKB solutions in vicinity
of the turning points, the corresponding Bohr-Sommerfeld quantization condition was obtained in \cite{quartic} which contains a non-perturbative in $\hbar$ term related to hyperasymptotics.

Here we use the results obtained in \cite{quartic} for the harmonic and quartic potentials to analyse the energies of  bound states  accounting also for higher order (up to $\hbar^4$)
WKB corrections. The emphasis is placed on the potential of an anharmonic oscillator $V(x) \sim x^4$, which has been extensively studied in quantum mechanics for many years (see, book \cite{Turbiner-book}
and the recent article \cite{ Babenko2025} with numerous references therein).
Since, to the best of our knowledge, there are no exact solutions to the double quartic problem (with the Hamiltonian $H=a^4\hat{p}^4+b^4x^4$) and to check
the WKB results, we determine the bound state energies numerically by applying the variational approach utilizing the universal Gaussian basis. In addition, it is always instructive to have
an exact solvable example, therefore, we find bound state energies and their wave functions for the exactly solvable problem of a square potential well.

One of our main findings is that in the case of quartic dispersion, although the wave functions of bound states in the classically forbidden region exponentially decay, they have nodes  in that
region  in all considered examples. This invalidates the oscillation theorem characteristic for the one-dimensional second-order Schrödinger equation. The latter states that the $n$th
eigenenergy bound-state wave function of the 1D Schrödinger equation  has exactly $n$ nodes within classically allowed region and no nodes in classically forbidden region.
Still it is observed that the oscillation theorem holds in the classically allowed region for  the fourth-order Schrödinger equation.

The paper is organized as follows. The WKB approximation and the Wentzel complex method are considered in Sec. \ref{sec:WKB} for the double quartic problem. In Sec.\ref{sec:numerical-analysis},
 bound state energies and wave functions are analysed using the variational method. Bound states for the square well potential are found in Sec.\ref{sec:potential-well}.  The results
are summarized in Sec.\ref{sec:conclusions}.

\section{Quantization condition for double quartic problem}
\label{sec:WKB}

The Hamiltonian of one-dimensional quasiparticles with quartic energy-momentum dispersion is given by
\begin{align}
H=a^4\hat{p}^4+V(x),
\label{Hamiltonian}
\end{align}
where $\hat{p}=-i\hbar \partial_x$ is the momentum operator, $a^4$ is parameter whose dimension is $v^4/W^3$, where $v$ is velocity and $W$ is energy
(compare with the energy dispersion in tetra-layer graphene where $E(p)=(v_Fp)^4/\gamma^3_1$ \cite{Koshino,Min,Jia}). We assume the potential $V(x)$ to be a continuous function which tends to infinity for $|x| \to \infty$ and has a single global minimum which ensures the existence of two real turning points.

To solve the stationary Schrödinger-like equation $H\psi=E\psi$ we use the WKB ansatz for the wave function
$\psi(x)=\exp[iS(x)/\hbar]$ and obtain the following equation:
\begin{align}
a^4\left[(S^{\prime})^4-6i\hbar S^{\prime\prime}(S^{\prime})^2-\hbar^2(4S' S'''+3(S'')^2)+i\hbar^3 S^{IV}\right]+V(x)=E,
\label{semiclassical-equation}
\end{align}
where primes denote derivatives with respect to $x$. Expanding $S$ in powers of $\hbar$, $S=\sum_{n=0}^\infty\hbar^n S_n$, one can get recursive relations for derivatives of
$S_n$ (for some low order recursive relations see Ref.\cite{quartic}). For example, in the zero order, we find the classical momentum
\begin{equation}
p(x)=S^{\prime}_0(x)=(E-V(x))^{1/4}/a,
\label{semiclassical-momentum}
\end{equation}
which gives $S_0(x)=\int^x dx(E-V(x))^{1/4}/a$. From recursive relations we find for $S^\prime_n$ the following expressions up to $n=4$:
\begin{align}
\label{second-order-correction}
&S^{\prime}_1=\frac{3i}{2}\left(\ln p(x)\right)^{\prime},\quad S^{\prime}_2=\frac{5}{4}\left(\frac{3(p')^2}{2p^3}-\frac{p''}{p^2}\right)= -\frac{5}{4}\left(\frac{(p')^2}{2p^3}
+\left(\frac{p'}{p^2}\right)'\right),\\
&S^{\prime}_3= \frac{5i}{8}\left(\frac{3(p')^2}{2p^4}-\frac{p''}{p^3}\right)^{\prime}, \qquad  S^{\prime}_4=\frac{35(p^{\prime\prime})^2}{256\,p^5}-\frac{21p^{IV}}{1024\,p^4}
+\left(\frac{75(p')^3}{256p^6}-\frac{75p'p''}{256p^5}+\frac{85p'''}{1024p^4}\right)^\prime.
\end{align}
Integrating the equation for $S_1'$, we obtain $S_1=i\ln p^{3/2}$.  Notice that terms in the expansion $S'=\sum_{n=0}^\infty\hbar^n S'_n$ with odd $n\ge3$ are total derivatives like in
the case of the Schrödinger equation and vanish after integrating over the closed contour in the complex plane encircling turning points \cite{Ito2021}. The total derivatives terms in $S'_n$
with even $n\ge2$ do not contribute either.

The higher order terms with $n\ge6$ can be also obtained recursively and expressed through $p(x)$ and its derivatives (for terms $S'_6,S'_8$, see Ref.\cite{Ito2021}). However,
the algebraic complexity of expressions increases rapidly with increasing $n$ and, at a certain stage, requires the use of a computer. In this paper, we restrict ourselves to the corrections up
to the 4th order. We now use the obtained results and apply the Wentzel complex method to determine the bound state energies for the system with the quartic dispersion and the quartic potential
$V(x)=b^4x^4$.

\begin{figure}
\centering
\includegraphics[scale=0.52]{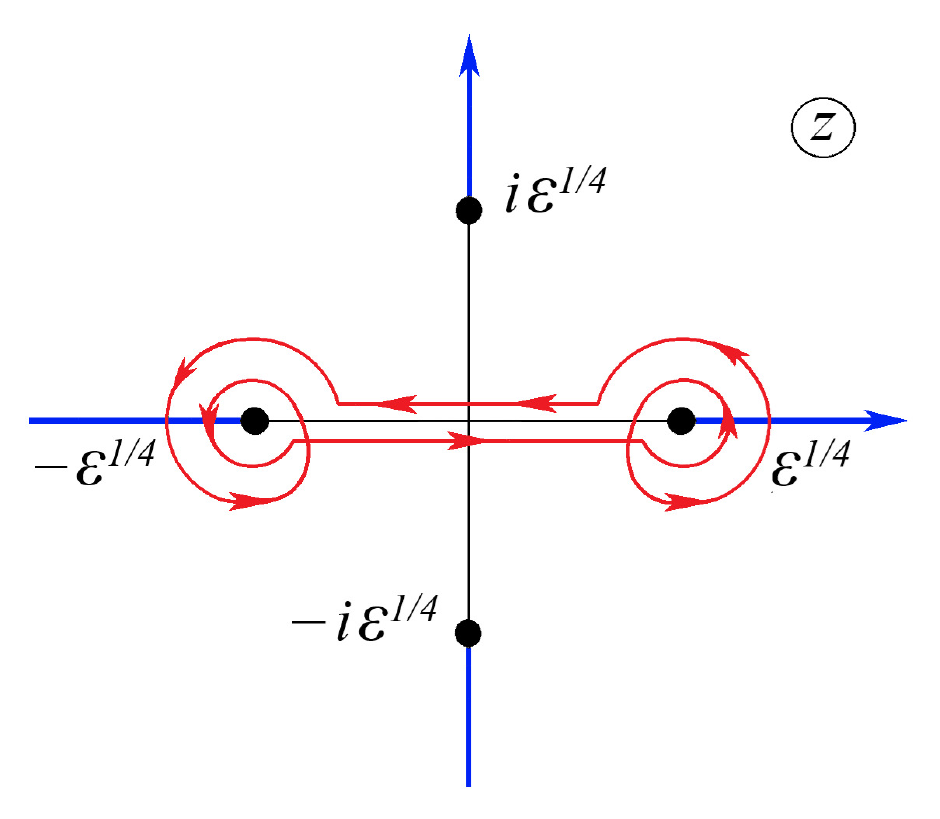}
\caption{Contour enclosing two turning points.}
\label{fig:contour}
\end{figure}

The Wentzel complex method uses the single-valuedness of a wave function $\psi(z)$ in the region of complex plane $z$ containing turning points. Indeed, on the real axis
the bound state wave function is a real function and must decrease in the classically forbidden region while oscillating in the classically allowed region producing nodes. To count for nodes due to
oscillations the argument principle should be used which then leads to the condition
\begin{align}
\oint_\gamma \frac{\psi'(z)}{\psi(z)}dz=\frac{i}{\hbar}\oint_\gamma S'(z)dz=2\pi i n,
\end{align}
where $n$ is an integer number and the contour $\gamma$ encloses the classical turning points anticlockwise in the complex $z$-plane. Since the semiclassical momentum is a fourth-root expression resulting in a four-sheeted Riemann surface of complex variable $z$, the contour should be chosen so that the integrand $S^{\prime}(z)$ return to its original value after going along the contour. Encompassing a turning point, the momentum acquires the phase  factor $e^{i\pi/2}$, therefore, one needs to go around each 
turning point twice,  as shown in Fig.\ref{fig:contour}, to return to the original value of the momentum. Since the integrand term $S^{\prime}_n(z)$, after calculating the derivatives, contains only integer powers of $z$ and $p(z)$, it is a single-valued function on the considered contour. In addition, the contour should be drawn close to the real axis avoiding possible zeros in the complex domain, as well as zeros in the classically forbidden domain, because we need 
to account for nodes of a bound state wavefunction on the real axis between turning points. Nodes in the classically forbidden region, as will be shown below, appear in eigenfunctions of the fourth-order differential operator 
for which, in contrast to the Schrödinger operator, the oscillation theorem does not hold in its standard formulation.

For quartic dispersion and in the fourth order in $\hbar$, we have
\begin{align}
\frac{1}{\hbar}\oint_\gamma S(z)dz=\frac{1}{\hbar}\oint_\gamma p(z)dz+\frac{3i}{2}\oint_\gamma \frac{p'(z)dz}{p(z)}
-\frac{5\hbar}{4} \oint_\gamma\frac{(p')^2}{2p^3}dz+\frac{7\hbar^3}{256} \oint_\gamma\left(\frac{5(p^{\prime\prime})^2}{p^5}-\frac{3p^{IV}}{4p^4}\right)dz =2\pi n.
\end{align}

To determine bound state energies for the double quartic problem (quartic dispersion and quartic potential)
\begin{equation}
\left(a^4\hbar^4\partial^4_x+b^4x^4\right)\psi(x)=E\psi(x),
\label{double-quartic-problem}
\end{equation}
it is convenient to use dimensionless quantities $z=x\sqrt{b/(a\hbar)}$, $\epsilon=E/(\hbar^2a^2b^2)$, and $p(z)=(\epsilon-z^4)^{1/4}$. Then we set the branch cuts to be the half-lines $(-\infty,-\epsilon^{1/4}]$, $[\epsilon^{1/4},\infty)$, $(-i\infty,-i\epsilon^{1/4}]$, and $[i\epsilon^{1/4},i\infty)$ so that $p(z)=(\epsilon-z^4)^{1/4}$
takes positive real values on the real axis between the turning points $-\epsilon^{1/4}$ and $\epsilon^{1/4}$ ($\epsilon>0$). Further, $p(z)$
is a single-valued function on the four-sheeted Riemann surface $p^4+z^4=\epsilon$ of genus 3. The closed contour $\gamma$ is 1-cycle on this surface which runs twice around
the turning point $\epsilon^{1/4}$, then twice around the point $-\epsilon^{1/4}$  as shown schematically in Fig.\ref{fig:contour}. Therefore,
calculating higher order WKB corrections we could omit terms with total derivatives which give zero.

For the double quartic problem, to determine higher order WKB corrections we need to calculate the contour integrals
\begin{align}
&\oint_\gamma p(z)dz= 2\int\limits_{-\epsilon^{1/4}}^{\epsilon^{1/4}}(\epsilon-y^4)^{1/4}dy= 2\frac{\sqrt{\epsilon}\Gamma^2(1/4)}{4\sqrt{\pi}},\\
&\oint_\gamma\frac{p'(z)dz}{p(z)}=\oint_\gamma\frac{d p^4(z)}{4p^4(z)} =2\pi i,
\end{align}
where we took into account in the last integral that each pole due to zeros of $p^4(z)$ is enclosed twice.
For corrections with $n\ge2$, after calculating derivatives $\partial_z^kp(z)$,  reducing the integer powers of $z$ which are multiples of $z^4$, and using $z^4=\epsilon-p^4$, we obtain
an expression which is given as a sum of terms with negative powers of $p(z)$. Representing negative powers of $p(z)$ as derivatives with respect to $\epsilon$,
we can write, e.g., the first term in the second order WKB correction (\ref{second-order-correction}) in the form of the Picard-Fuchs differential operator acting on the period integrals of $p(z)$
\begin{align}
\oint_\gamma\frac{(p')^2}{p^3}dz=\oint_\gamma\frac{z^6dz}{p^9(z)}=\oint_\gamma\left(\frac{\epsilon z^2}{p^9(z)}-\frac{ z^2}{p^5(z)}\right)dz=
\left(\frac{16}{5}\epsilon\partial^2_\epsilon+4\partial_\epsilon\right)\oint_\gamma\frac{z^2dz}{p(z)}= 2\frac{6\Gamma^2(3/4)}{5\sqrt{\pi}}\frac{1}{\sqrt{\epsilon}}.
\end{align}
Similarly, for the fourth order correction integral, we get
\begin{align}
&\frac{7}{256}\oint_\gamma\left(\frac{5(p^{\prime\prime})^2}{p^5}-\frac{3p^{IV}}{4p^4}\right)dz=\frac{7}{256}\oint_\gamma\left(\frac{135\epsilon}{2p^{11}}-\frac{1125\epsilon^2}{4p^{15}}
+\frac{873\epsilon^3}{4p^{19}}\right)dz\nonumber\\
&=\left[\frac{45}{8}\epsilon\partial^3_\epsilon+\frac{375}{44}\epsilon^2\partial^4_\epsilon+\frac{97}{55}\epsilon^3\partial^5_\epsilon\right]\oint_\gamma p(z)dz
=-2\frac{3\Gamma^2(1/4)}{128\sqrt{\pi}\epsilon^{3/2}}.
\end{align}
Finally, we obtain the following quantization condition which defines the WKB energy levels of the double quartic problem:
\begin{align}
\frac{\Gamma^2(1/4)\sqrt{\epsilon}}{4\sqrt{\pi}}-\frac{3\Gamma^2(3/4)}{4\sqrt{\pi}}\frac{1}{\sqrt{\epsilon}}-\frac{3\Gamma^2(1/4)}{128\sqrt{\pi}\epsilon^{3/2}}
-2(-1)^n\arctan\left[\frac{1}{2}\exp\left(-\frac{\Gamma^2(1/4)\sqrt{\epsilon}}{4\sqrt{\pi}}\right)\right]=\pi\left(n+\frac{1}{2}\right),\qquad n=0,1,\dots\,,
\label{energies-with-hyperasymptotics}
\end{align}
where the last term on the left-hand side of the above equation accounts for the non-perturbative contribution due to hyperasymptotics derived in Ref.\cite{quartic}.

\begin{table}[ht]
\begin{tabular}{|l|l|l|l|l|l|l|l|}
\hline
$\varepsilon$ \textbackslash \,\,$n$ & 0 & 1 & 2 & 3 & 4 & 5 & 6 \\
\hline
$\varepsilon_{4thcorr}$ &1.4175 & 7.1539 & 18.6332 & 35.8581 & 58.8260 & 87.5364 & 121.9891\\
\hline
$\varepsilon_{4hyper}$ & 1.5151 & 7.1345 & 18.6348 & 35.8580 & 58.8260 & 87.5364 & 121.9891\\
\hline
$\varepsilon_{var}$ & 1.3967& 7.1315 & 18.6404 & 35.8590 & 58.8263 &87.5365&121.9892\\
\hline
\end{tabular}
\caption{Numerical values of the bound state energy $\varepsilon$ for the one-dimensional system with quartic dispersion and quartic potential for $n=0,1,\dots,6$ taking
into account the WKB corrections of the second  and fourth order $\varepsilon_{4thcorr}$,  as well as the non-perturbative correction due to hyperasymptotics. Bound-state
energies obtained using the variational approach in Sec.\ref{sec:numerical-analysis} are shown in the last row.}
\label{table-of-energies}
\end{table}

Numerical results of computation of the WKB energies for the double quartic system defined by Eq.(\ref{energies-with-hyperasymptotics}) are presented in Table \ref{table-of-energies} for the states with $n=0,1,...,6$. The second row shows the WKB energies with the second and fourth order WKB corrections taken into account. To quantify the role of the non-perturbative contribution, we include in the third row the non-perturbative in $\hbar$ contribution given by the last term on the left-hand side of Eq.(\ref{energies-with-hyperasymptotics}). The last row presents variational estimates which are derived in the next section.

Comparing these numerical values, we conclude that both WKB corrections and the non-perturbative contribution are indeed relevant for the lowest bound state energies. Note also that the non-perturbative correction due to hyperasymptotics in the Bohr-Sommerfeld quantization condition is always smaller than $\pi/2$ in accordance with similar results obtained for the Schrödinger equation \cite{Turbiner}. It can be seen also that the deviation of $\varepsilon_{4hyper}$ from the variational value $\varepsilon_{var}$ is around 8 per cent for the lowest state and then quickly decreases.

\section{Variational method and nodal structure of bound-state wave functions}
\label{sec:numerical-analysis}

To analyse the properties of bound-state wave functions and check the validity and accuracy of the results obtained in the previous section, we apply here the variational method with the universal Gaussian basis \cite{R4} for systems with quartic energy-momentum dispersion and the harmonic $V\left(x\right) \sim x^{2}$ and quartic $V\left(x\right) \sim x^{4}$ potentials.

In the universal Gaussian basis and dimensionless variables,  the trial wave functions are represented as a series expansion  of basic Gaussian functions
\begin{align}
\psi\!\left(x\right)\!=\!\sum_{k=1}^{\mathcal{K}}C_{k}\chi_{k}\left(x\right)\!\equiv\!\sum_{k=1}^{\mathcal{K}}\!C_{k}\exp\!\left(\!-a_{k}\left(x\!-\!x_{k}\right)^{2}\right)\!,
\label{E8}
\end{align}
where $C_{k}$, $a_{k}$, and $x_k$ are variational parameters. Note that this basis enables one to consider both symmetric (even parity) and antisymmetric (odd parity) states within the same variational procedure, while if $x_{k}=0$, then we need to deal with symmetric states and antisymmetric states separately. To find $C_{k}$, it is suitable to use the Galerkin method and solve the algebraic system of equations
\begin{align}\label{E3}
\sum_{k=1}^{\mathcal{K}}C_{k}\left\langle\varphi_{n}\right|H-E\left|\varphi_{k}\right\rangle=0, \,\,\, n=1,...,\mathcal{K},
\end{align}
which is equivalent to the Schr\"{o}dinger equation for $\mathcal{K}\rightarrow \infty$. Parameters $x_k$ and $a_{k}$ can be found within different schemes of variational procedures or chosen randomly using the stochastic variational method \cite{R2,R3}. For the matrix elements of powers of $x$ and $\hat{p}$ (where $\hat{p}=-i\partial_x$), we have
\begin{align*}
\left\langle\chi_{n}\right|\hat{p}^{2}\left|\chi_{k}\right\rangle=\frac{2\sqrt{\pi}\,a_{k}a_{n}}{u_{nk}^{3/2}}\left(1-2\omega_{nk}\right)\exp\left(-\omega_{nk}\right),\quad
\left\langle\chi_{n}\right|\hat{p}^{4}\left|\chi_{k}\right\rangle=\frac{4\sqrt{\pi}\,a_{k}^{2}a_{n}^{2}}{u_{nk}^{5/2}}\left(3-12\omega_{nk}+4\omega_{nk}^{2}\right)\exp\left(-\omega_{nk}\right),
\end{align*}
\begin{align}
\left\langle\chi_{n}\right|x^{2}\left|\chi_{k}\right\rangle=\frac{\sqrt{\pi}}{2u_{nk}^{3/2}}\left(1+2\lambda_{nk}^{2} u_{nk}\right)\exp\left(-\omega_{nk}\right)\!,\,\,\,\,\,\,\left\langle\chi_{n}\right|x^{4}\left|\chi_{k}\right\rangle=\frac{3\sqrt{\pi}}{4u_{nk}^{5/2}}\left(1+4\lambda_{nk}^{2}u_{nk}+\frac{4}{3}\lambda_{nk}^{4}u_{nk}^{2}\right)\exp\left(-\omega_{nk}\right),
\label{E9}
\end{align}
where
\begin{align*}
\lambda_{nk}\equiv\frac{a_{n}x_{n}+a_{k}x_{k}}{u_{nk}}, \quad \omega_{nk}\equiv\frac{a_{n}a_{k}}{u_{nk}}\left(x_{n}\!-\!x_{k}\right)^{2},\quad u_{nk}=a_{n}+a_{k}.
\end{align*}
Finally, for the normalization of matrix elements, we find
\begin{align*}
\left\langle\chi_{n}\right|\left.\chi_{k}\right\rangle=\frac{\sqrt{\pi}}{\sqrt{u_{nk}}}\exp\left(-\omega_{nk}\right).
\end{align*}

In Table~II, we present the variational estimations for a few lowest energy levels obtained using the universal basis (\ref{E8}) for the Hamiltonian $H=\hat{p}^{4}+x^{4}$. These energy levels lies higher then those obtained for the Hamiltonian $H=\hat{p}^{4}+x^{2}$ and the convergence of the consequent variational estimations at $\mathcal{K}\rightarrow\infty$ is noticeably more slow. This may be connected with the fact that the wave functions of the problem with quartic dispersion have nontrivial asymptotics containing oscillations in the classically forbidden regions \cite{quartic}. This means that nontrivial kinetic energy $\sim \hat{p}^{4}$, generally speaking, violates the well-known oscillation theorem for the quadratic dispersion $E \sim \hat{p}^{2}$ (see, e.g, \cite{Morse}) which relates the number of nodes of a bound state wave function with its quantum number $n$. According to \cite{quartic}, eigenfunctions of the Hamiltonian $H=\hat{p}^{4}+x^{4}$ have the following behaviour in the classically forbidden region at $x\rightarrow +\infty$:
\begin{align}\label{E13}
\psi_n\left(x\right) \rightarrow C_n\cdot\exp\left(-\frac{x^{2}}{2\sqrt{2}}\right)\cos\left(\frac{x^{2}}{2\sqrt{2}}+\theta_n\right),
\end{align}
where $C_n$ and $\theta_n$ are constants. At $x\rightarrow -\infty$, symmetric states have the same asymptotics (22), while antisymmetric states change their overall sign. This asymptotics is valid for the ground state too and leads to an infinite number of nodes of the ground state wave function. In addition, such a nontrivial oscillating asymptotics of wave functions requires a large number of basic functions to reproduce accurately the wave functions under study in the whole interval of $x$.

Finally, we wish to stress that all presented numbers in Table~II are only the variational estimations from above, and we do not present here the ``exact'' values which, to the best of our knowledge, are not available in the literature. Still, due to the convergence of our results as we extend the number of basis functions, we think that our values found for $\mathcal{K}=20$ provide the correct results with accuracy in the last decimal digit.

\begin{table*}[tbp] \centering
\begin{tabular}
[c]{|c|c|c|c|c|c|c|c|c|c|c|c|}
\hline  $n$ & $\mathcal{K}=1$ & $\mathcal{K}=2$ & $\mathcal{K}=3$ & $\mathcal{K}=4$ & $\mathcal{K}=5$ & $\mathcal{K}=6$ & $\mathcal{K}=7$ & $\mathcal{K}=8$ & $\mathcal{K}=12$ & $\mathcal{K}=16$ & $\mathcal{K}=20$
\\\hline
$\,0\,$ & 1.500000 & 1.429335 & 1.396878 & 1.396797 & 1.396735 & 1.396728 & 1.396728 & 1.396728 & 1.396728 & 1.396728 & 1.396728
\\\hline
$\,1\,$ &  & 7.230771 & 7.182758 & 7.136187 & 7.132242 & 7.131574 & 7.131535 & 7.131529 & 7.131529 & 7.131529 & 7.131529
\\\hline
$\,2\,$ &  &  & 18.738913 & 18.664837 & 18.643336 & 18.640507 & 18.640434 & 18.640383 & 18.640383 & 18.640383 & 18.640383
\\\hline
$\,3\,$ &  &  &  & 35.902068 & 35.879689 & 35.866030 & 35.859094 & 35.859036 & 35.859024 & 35.859024 & 35.859024
\\\hline
$\,4\,$ &  &  &  &  & 58.956623 & 58.846621 & 58.827356 & 58.826379 & 58.826330 & 58.826330 & 58.826330
\\\hline
$\,5\,$ &  &  &  &  &  & 88.826155 & 87.545067 & 87.536934 & 87.536541 & 87.536528 & 87.536528
\\\hline
$\,6\,$ &  &  &  &  &  &  & 122.042196 & 121.998877 & 121.989268 & 121.989198 & 121.989193
\\\hline
\end{tabular}
\caption{Successive approximations for the energy levels for the Hamiltonian $H=\hat{p}^{4}+x^{4}$ calculated within the variational procedure using the variational basis (\ref{E8}) for different number $\mathcal{K}$ of the Gaussian functions. Empty cells in the left lower part of the table are present because to determine $\mathcal{K}$ energy levels one should use not less than $\mathcal{K}$ basis functions in the variational ansatz.}%
\label{Tab:1}
\end{table*}%

\begin{figure}
\centering
\includegraphics[scale=0.38]{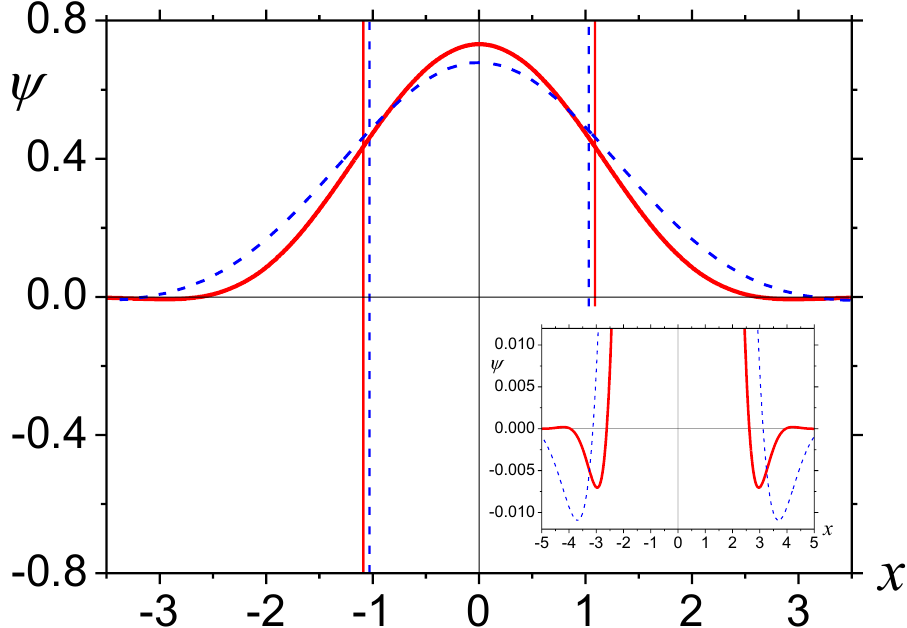}
\includegraphics[scale=0.38]{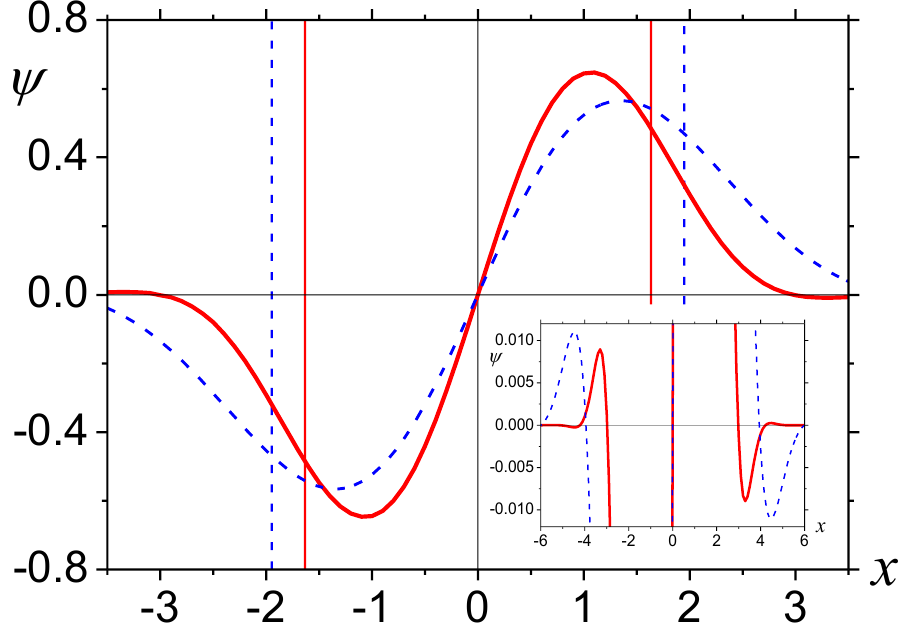}
\includegraphics[scale=0.37]{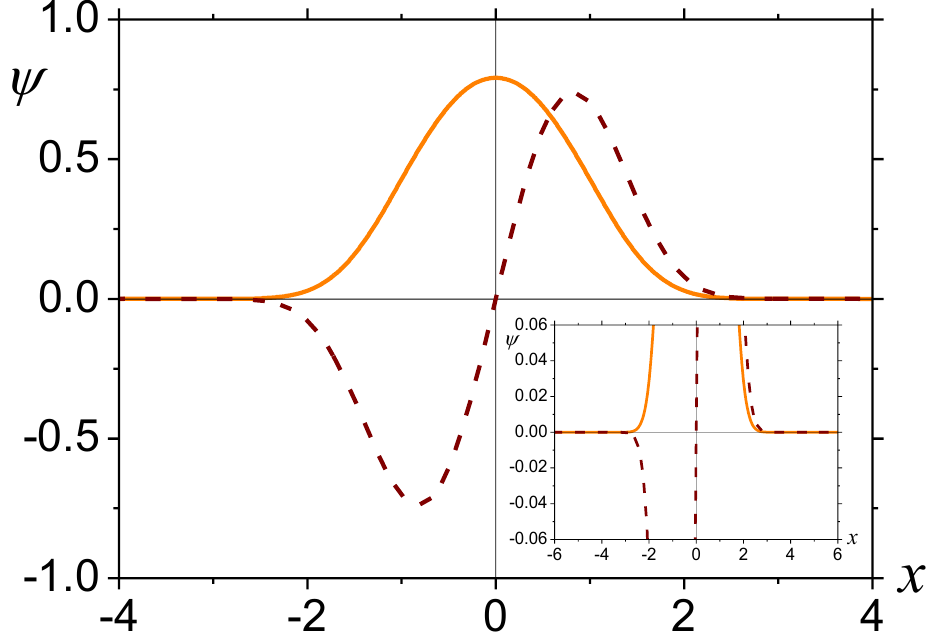}
\caption{Left panel: The ground state wave function for the $H=\hat{p}^{4}+x^{4}$ (red solid line) and $H=\hat{p}^{4}+x^{2}$ Hamiltonians (blue dashed line). Middle panel: The wave functions of the first excited state for the $H=\hat{p}^{4}+x^{4}$ (red solid line) and $H=\hat{p}^{4}+x^{2}$ Hamiltonians (blue dashed line).  Vertical red solid and blue dashed lines in the left and middle panels show boundaries of the classically allowed region for the ground and first excited states. Right panel: The wave function of the ground state (orange solid line) and the first excited state (brown dashed line) for the quartic oscillator with the Hamiltonian $H=\hat{p}^{2}+x^{4}$.}
\label{fig:wave-functions-1}
\end{figure}

Let us consider now the wave functions of the bound states. Using Eq.(\ref{E8}) and the determined coefficients $C_{k}$ and parameters $a_{k}$, $x_{k}$, we depict the bound state
wave functions of the ground and first excited states for the Hamiltonian $H=\hat{p}^4+x^4$ in the left and middle panels of Fig.\ref{fig:wave-functions-1} (red solid lines). For comparison, we show in the same figure the wave functions for the harmonic potential $V\left(x\right)=x^{2}$ (blue dashed lines). Energies of the ground and first excited states for the Hamiltonian $H=\hat{p}^4+x^4$ are $E=1.3967$
and $E=7.1315$, respectively. For the Hamiltonian with quadratic potential $H=\hat{p}^4+x^2$, the corresponding energies are lower and equal $E=1.0603$
and $E=3.7996$. Vertical red solid and blue dashed lines in the left and middle panels show boundaries of the classically allowed region for the ground and first excited states, respectively.
While for the ground state of the Hamiltonian $H=\hat{p}^4+x^4$ the classically allowed region is defined by $-1.0871<x<1.0871$ for the ground state, it is given by $-1.6341<x<1.6341$ for
the first excited state. As to the classically allowed region for the Hamiltonian $H=\hat{p}^4+x^2$, it is $-1.0297<x<1.0297$ and $-1.94927<x<1.94927$ for the ground and first excited states,
respectively. The wave functions are normalized to the unity, i.e., $\int \left|\psi\left(x\right)\right|^{2}dx=1$.
Oscillations of the wave functions of the ground and first excited states for the potential $V(x)=x^4$ in the classically forbidden region are shown in insets of all panels of Fig.\ref{fig:wave-functions-1}. It is instructive to compare the wave functions of the ground and first excited states of the double quartic problem with the familiar case of the quadratic energy-momentum dispersion. The wave functions of the later problem are plotted in the right panel of Fig.\ref{fig:wave-functions-1}.

\begin{figure}
\centering
  \includegraphics[scale=0.36]{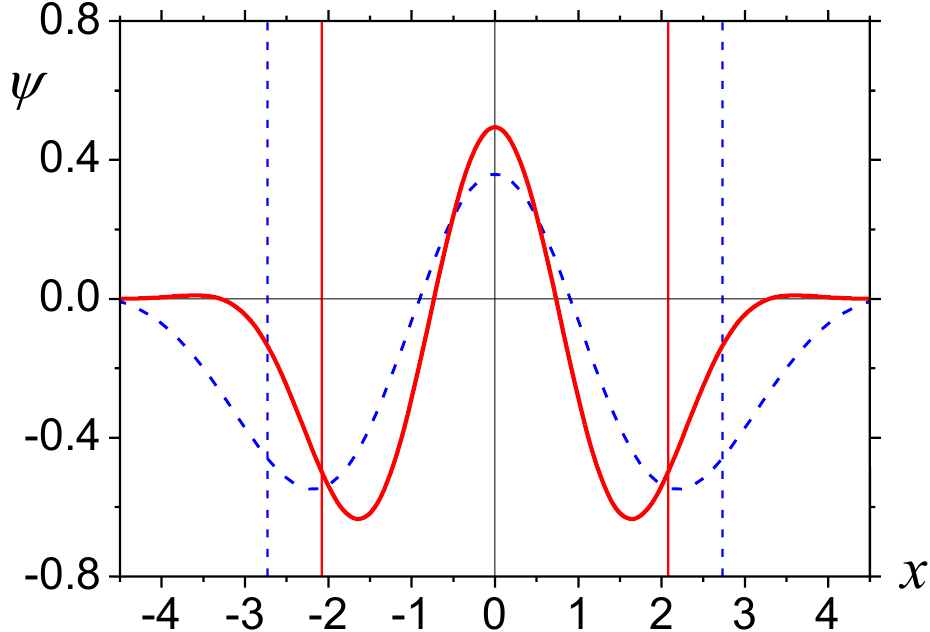}\hspace{5mm}
\includegraphics[scale=0.36]{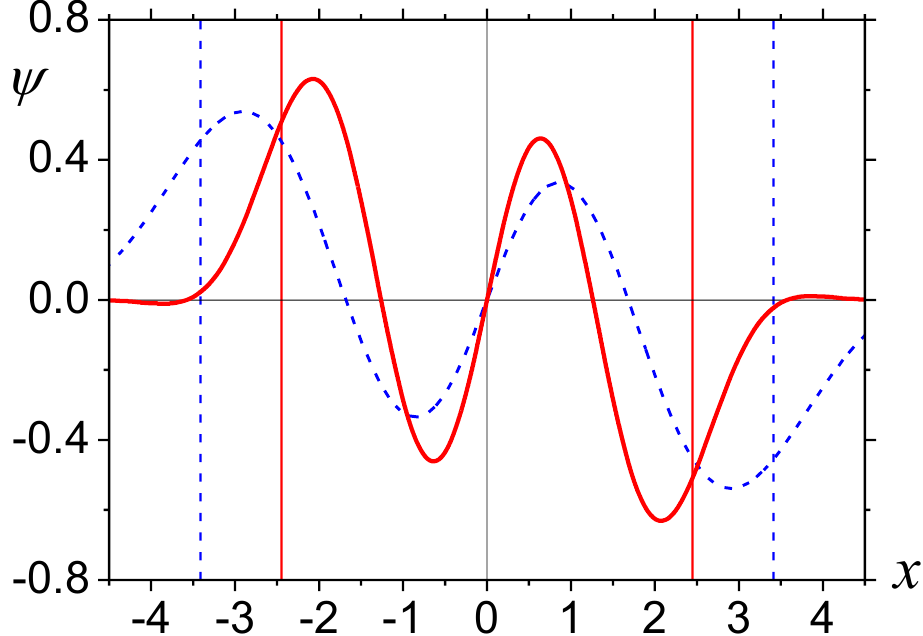}
\includegraphics[scale=0.36]{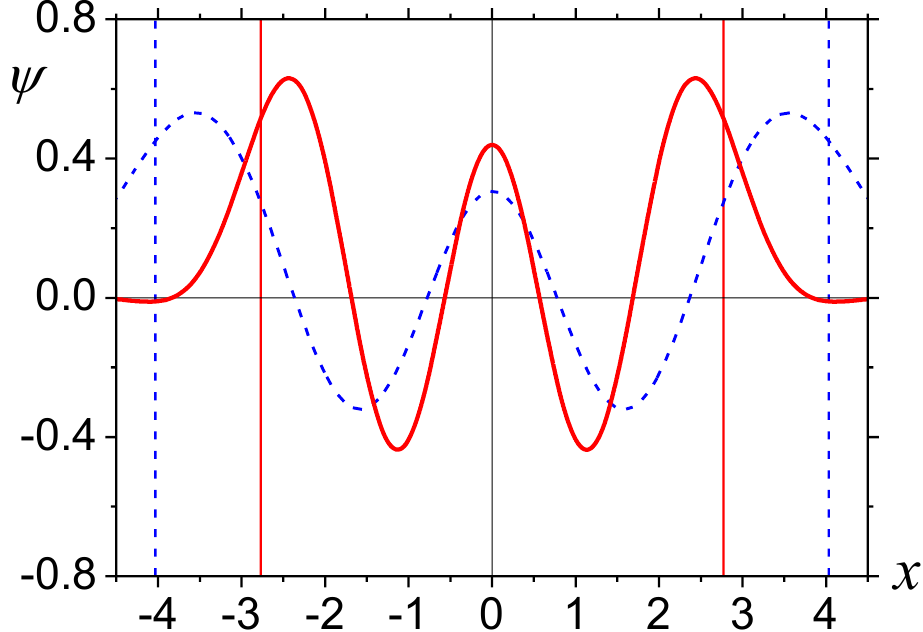}
\caption{The wave functions of the second (left panel), third (middle panel), and fourth excited state (right panel) for the potentials $V(x)=x^4$ (red solid lines) and $V(x)=x^2$ (blue dashed lines). Vertical red solid and blue dashed lines show boundaries of the corresponding classically allowed regions.}
\label{fig:wave-functions-2}
\end{figure}

The wave functions of the 2nd, 3rd, and 4th excited states are shown in Fig.\ref{fig:wave-functions-2}, where vertical red solid and blue dashed lines show boundaries of the classically allowed region for the potentials $V(x)=x^4$ and $V(x)=x^2$, respectively. Note that the number of nodes equals $n$ for the $n$th energy level in the classically allowed region. Certainly, the presence of nodes in the wave functions in the classically forbidden region clearly distinguishes the bound state problem for quasiparticles with quartic dispersion from that with the conventional quadratic dispersion. This result is definitely worth checking and illustrating by analysing an exactly solvable problem. As such a problem, we consider in the next section a square well potential, which has not been discussed in detail in the literature. We determine the bound state energy spectrum and the wave functions which show explicitly the presence of oscillations in the classically forbidden region, thus, confirming the violation of the oscillation theorem for systems with quartic dispersion.

\section{Spectrum and wave function nodes for square well potential}
\label{sec:potential-well}

Let us determine bound state energies and the corresponding wave functions for an exactly solvable problem of  quasiparticles with quartic dispersion in a square well potential. We consider the one-dimensional Hamiltonian (\ref{Hamiltonian})
with potential
\begin{align}
V(x)=\left\{\begin{array}{c}V_0>0,\quad \mbox{for}\quad |x|>L,\\
0,\qquad \mbox{for}\quad -L<x<L.\end{array}\right.
\end{align}
Energy levels and wave functions of bound states are defined by the equation
\begin{align}
\psi^{IV}(x)+\frac{V(x)-E}{\hbar^4a^4}\psi(x)=0.
\end{align}
We seek for bound state solutions which can exist for energies $0<E<V_0$. In the regions $x<-L$ and $x>L$, the corresponding solutions
decreasing for $x\to\mp\infty$, respectively, are
\begin{align}
&\psi_I(x)=A_1 e^{\kappa_1x}+A_2 e^{\kappa_2x},\qquad x<-L,\\
&\psi_{III}(x)=B_1 e^{-\kappa_1x}+B_2 e^{-\kappa_2x},\qquad x>L,
\end{align}
where
\begin{align}
\kappa_{1,2}=\frac{1\pm i}{\sqrt{2}}\frac{(V_0-E)^{1/4}}{\hbar a}.
\end{align}
In the region $-L<x<L$, we have four independent solutions
\begin{align}
\psi_{II}(x)=\sum\limits_{k=1}^4C_k e^{i p_kx},\quad p_k=e^{\frac{i\pi k}{2}}\frac{E^{1/4}}{\hbar a},\qquad -L<x<L.
\end{align}
Matching these functions and their three derivatives at $x=-L$ and $x=L$ we get the system of eight homogeneous linear equations for eight constants
$A_1,A_2,B_1,B_2,C_1,C_2,C_3,C_4$. By setting the determinant of this system equal to zero, we obtain the characteristic equation for the energy levels.

Actually, it is quite helpful in the case under consideration to simplify our task by using the symmetry of the Hamiltonian under the inversion transformation $x\to-x$ which separates all eigenstates in states of either positive or negative parity. Then it suffices to find solutions
only for positive $x$ using the basis of real functions. We begin our analysis with states of positive parity. The corresponding solution $\psi_{III}$ finite at $x\to+\infty$ can be written in the form
\begin{align}
\psi_{III}(x)= B_1 e^{-k x}\cos(k x)+B_2 e^{-k x}\mbox{sgn}(x)\sin(kx),\qquad k=\frac{(V_0-E)^{1/4}}{\sqrt{2}\hbar a},\qquad x>L.
\label{phi3-pp}
\end{align}
The general solution $\psi_{II}$ of positive parity in the classically allowed region $-L<x<L$ is
\begin{align}
\psi_{II}(x)=C_1\cos(p x)+C_2\cosh(p x),\qquad p=\frac{E^{1/4}}{\hbar a}.
\end{align}
The wave function $\psi_I$ in the classically forbidden region $x<-L$ is obtained by changing the sign of $x$ in $\psi_{III}(x)$. As mentioned above, in view of the inversion symmetry, it suffices to match solutions $\psi_{II}(x)$ and $\psi_{III}(x)$ and their first, second, and third derivatives at $x=L$. Then we get a system of four homogeneous linear equations
for four real constants $C_1,C_2,B_1,B_2$ with the matrix
\begin{align}
M=\left(\begin{array}{cccc}\cos(p L)&\cosh(p L)&-e^{-k L}\cos(k L)&-e^{-k L}\sin(k L)\\
-p\sin(p L)& p\sinh(p L)&k e^{-k L}(\cos(k L)+\sin(k L))&k e^{-k L}(-\cos(k L)+\sin(k L))\\
 -p^2\cos(p L)& p^2\cosh(p L)&-2k^2e^{-k L}\sin(k L)&2k^2e^{-k L}\cos(k L)\\
 p^3\sin(p L)&p^3\sinh(p L)&2k^3 e^{-k L}(-\cos(k L)+\sin(k L))&-2k^3 e^{-k L}(\cos(k L)+\sin(k L))\end{array}\right).
\end{align}
Equating the determinant of this matrix to zero, we obtain the following equation for energies of positive parity states in the potential well:
\begin{align}
2 \sqrt{2} k^3 p + (k^4 - 2 k^2 p^2 - p^4) \tan( p L) + (k^4 + 2 k^2 p^2 - p^4 - 2 \sqrt{2} k p^3 \tan( p L)) \tanh( p L)=0,
\label{eq:positive_parity}
\end{align}
where
\begin{align}
p=\epsilon^{1/4},\qquad k=(v-\epsilon)^{1/4}, \qquad \epsilon=\frac{E}{\hbar^4a^4},\qquad v=\frac{V_0}{\hbar^4a^4}
\label{def:p,k,epsilon,v}
\end{align}
[note that we redefined $k$ compared to Eq. (\ref{phi3-pp}) omitting $1/\sqrt{2}$].

Equation (\ref{eq:positive_parity}) can be written in another form
\begin{align}
e^{-2p L}\frac{p^2-\sqrt{2}p k+k^2}{p^2+\sqrt{2}p k+k^2}= -\frac{B\tan(p L)+A}{A\tan(p L)-B},\quad A=p^2-\sqrt{2}p k-k^2,\quad B=p^2+\sqrt{2}p k-k^2.
\label{positive_parity}
\end{align}

For negative parity states, general solutions are
\begin{align}
&\psi_{II}(x)=C_1\sin(p x)+C_2\sinh(p x),\qquad -L<x<L,\\
&\psi_{III}(x)= B_1 e^{-k x}{\rm sgn}(x)\cos(k x)+B_2 e^{-k x}\sin(k x),\qquad x>L.
\end{align}
Matching solutions $\psi_{II}(x)$ and $\psi_{III}(x)$ and their three derivatives at the point $x=L$ we get a system of four homogeneous linear equations
for four real constants $C_1,C_2,B_1,B_2$ with the matrix
\begin{align}
M=\left(\begin{array}{cccc}\sin(p L)&\sinh(p L)&-e^{-k L}\cos(k L)&-e^{-k L}\sin(k L)\\
p\cos(p L)& p\cosh(p L)&k e^{-k L}(\cos(k L)+\sin(k L))&k e^{-k L}(-\cos(k L)+\sin(k L))\\
 -p^2\sin(p L)& p^2\sinh(p L)&-2k^2e^{-k L}\sin(k L)&2k^2e^{-k L}\cos(k L)\\
- p^3\cos(p L)&p^3\cosh(p L)&2k^3 e^{-k L}(-\cos(k L)+\sin(k L))&-2k^3 e^{-k L}(\cos(k L)+\sin(k L))\end{array}\right).
\end{align}
Equating the determinant of this matrix to zero we obtain the eigenvalue equation
\begin{align}
\left[2\sqrt{2}k p^3 + (k^4 + 2 k^2 p^2 - p^4) \tan( p L) \right]\cosh(p L)  - \left[k^4 - 2 k^2 p^2 - p^4 - 2 \sqrt{2}k^3 p \tan( p L))\right]\sinh(p L)=0.
\label{eq:negative_parity}
\end{align}

\begin{figure}
\centering
\includegraphics[scale=0.22]{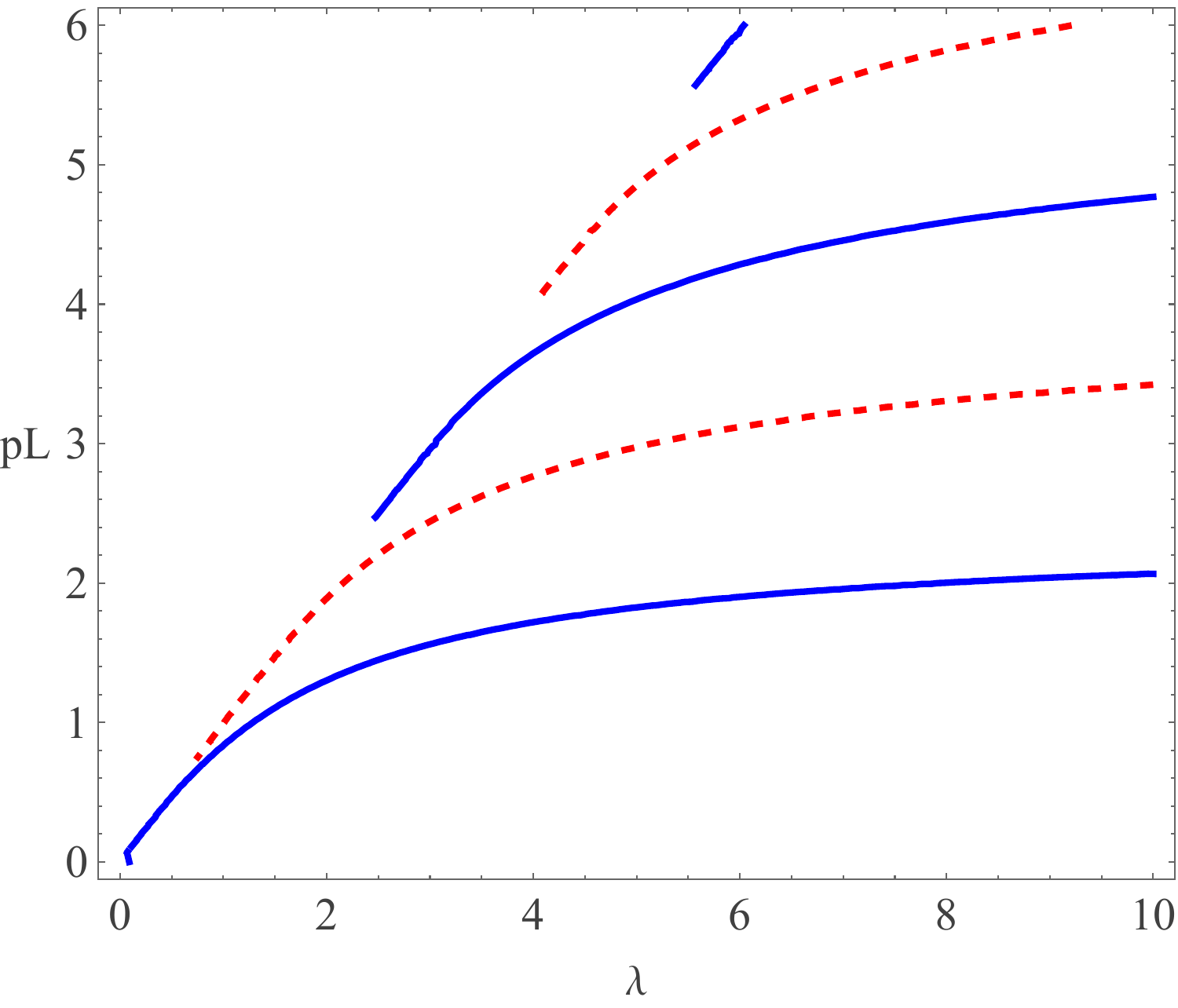}
\includegraphics[scale=0.22]{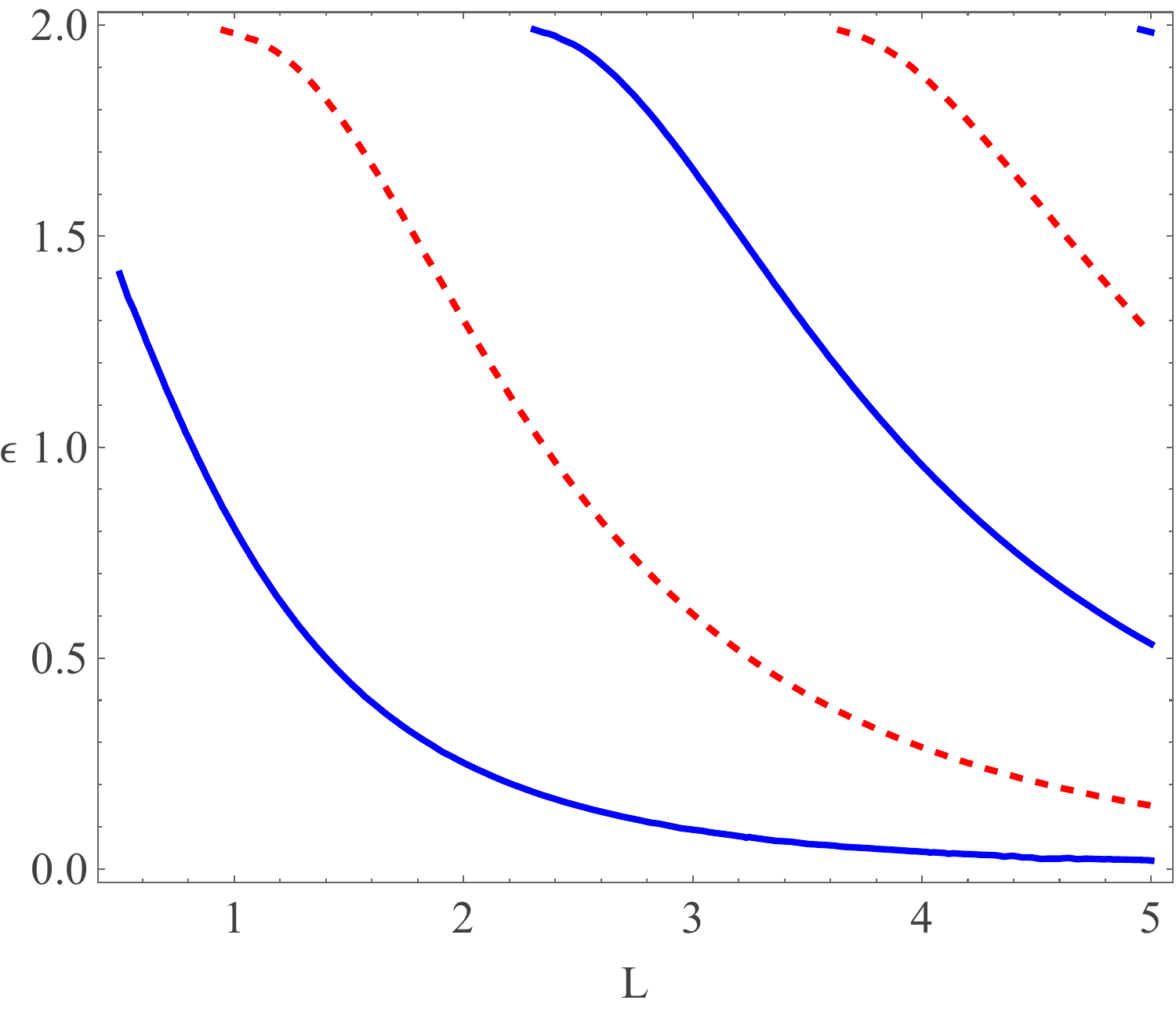}
\includegraphics[scale=0.18]{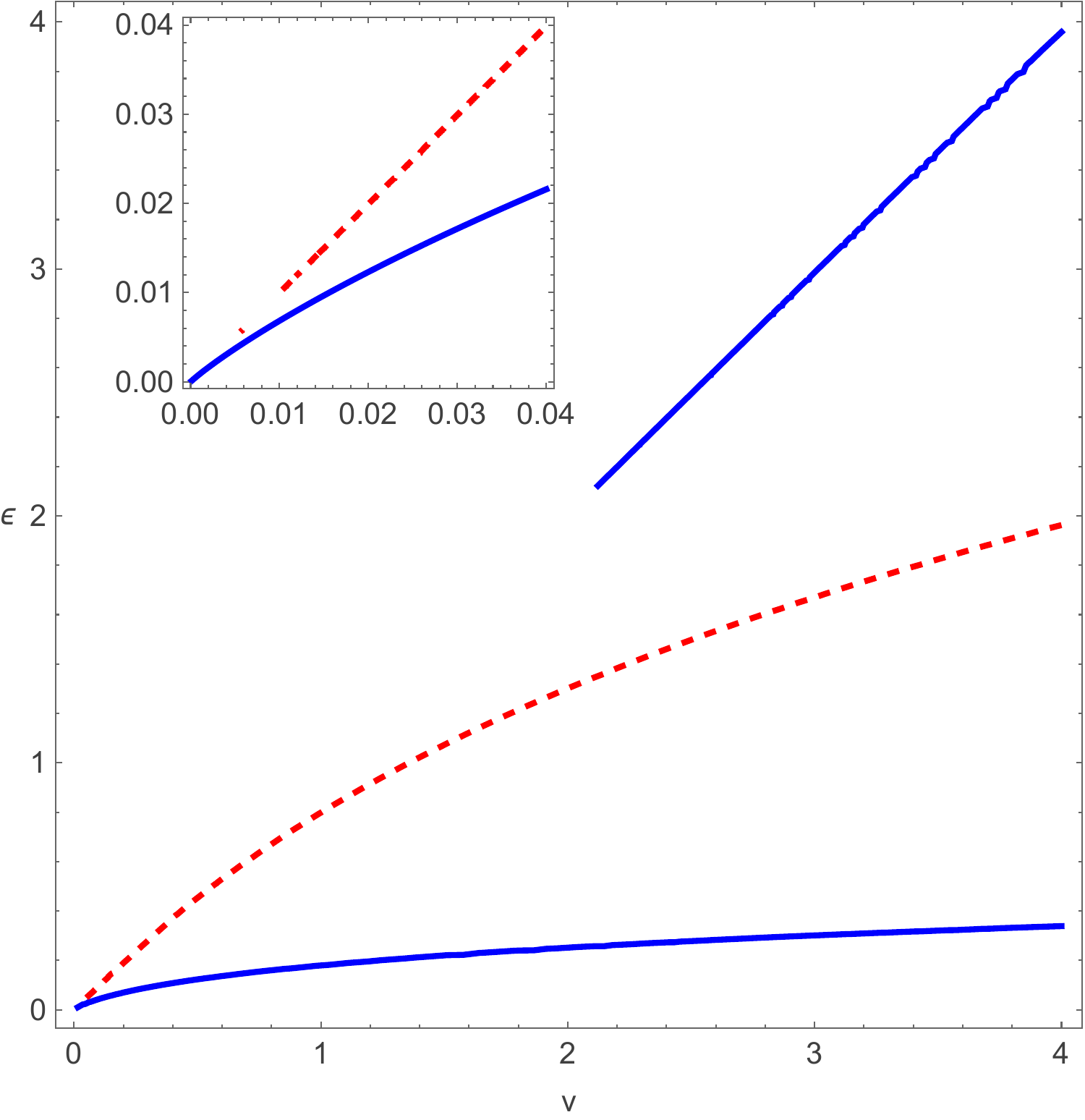}
\caption{Left panel: Energy levels $p L$ for positive and negative parity bound states as a function of $\lambda=v^{1/4}L$ (blue and red lines, respectively). Middle panel: energy levels as a function of the well width $L$ for the fixed well depth $v=2$. Right panel: energy levels as a function of the well depth $v$ for the fixed well width $L=1$.}
\label{fig:pp+np-lambda}
\end{figure}
The last equation can be written in the form
\begin{align}
e^{-2p L}\frac{p^2-\sqrt{2}p k+k^2}{p^2+\sqrt{2}p k+k^2}= -\frac{A\tan(p L)-B}{B\tan(p L)+A},\quad A=p^2-\sqrt{2}p k-k^2,\quad B=p^2+\sqrt{2}p k-k^2,
\label{negative_parity}
\end{align}
where $p$ and $k$ are defined in Eq.(\ref{def:p,k,epsilon,v}).

Equation (\ref{negative_parity}) agrees with the corresponding equation for negative parity states in Example 4 of Ref.\cite{Behncke} (modulo some misprints in the expressions for $A$ and $B$). The behaviour of energy levels given by Eqs.(\ref{positive_parity}) and (\ref{negative_parity}) as a function of $\lambda=v^{1/4}L$ is shown in the left panel of Fig.\ref{fig:pp+np-lambda}. The energies of states with negative parity lie between neighbouring states of positive parity and the ground state has positive parity.

Note that in the limit $v\to \infty$ (infinite square well), Eqs.(\ref{eq:positive_parity}) and (\ref{eq:negative_parity}) reduce in view of $k \to \infty$ to the following simple equation:
\begin{align}
\tan(p L)= \mp\tanh(p L)
\label{infinite-well}
  \end{align}
for states of positive and negative parity, respectively.  In contrast to the Schrödinger equation in the infinite well limit, Eq.(\ref{infinite-well}) is transcendental and corresponds to the infinite well problem with the boundary conditions $\psi(\pm L)=0$ and $\psi'(\pm L)=0$. As in the case of Schrödinger's equation, the energy levels in an infinite well potential are higher than the corresponding values for a finite well potential because the wave function in the latter case penetrates into
a classically forbidden region. Further, the wave function of the ground state is nodeless. In fact, for systems with quartic dispersion, it was proved \cite{Leighton,Rayleigh} that the node theorem is valid for infinite well $v \to \infty$. However, the wave function for a finite potential well problem has infinite number of oscillations in the classically forbidden region confirming our conclusion made in Sec.\ref{sec:numerical-analysis}. It is worth mentioning also that, for a finite width barrier, oscillations of wave functions may lead to the appearance of oscillating tunneling current \cite{Levitov,Sushkov}.

\begin{figure}
\centering
  \includegraphics[scale=0.44]{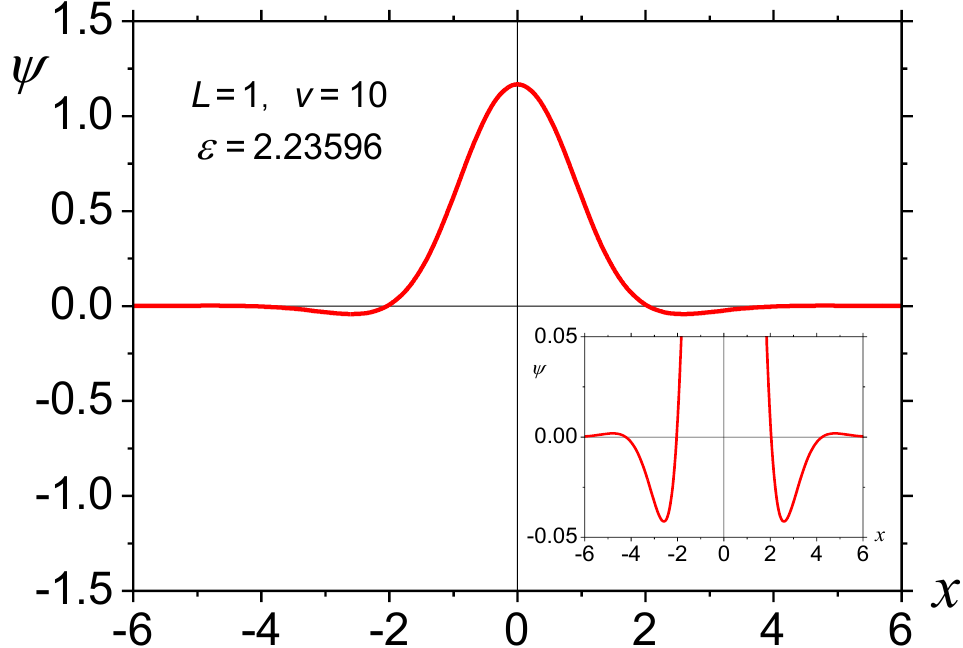}\hspace{5mm}
\includegraphics[scale=0.44]{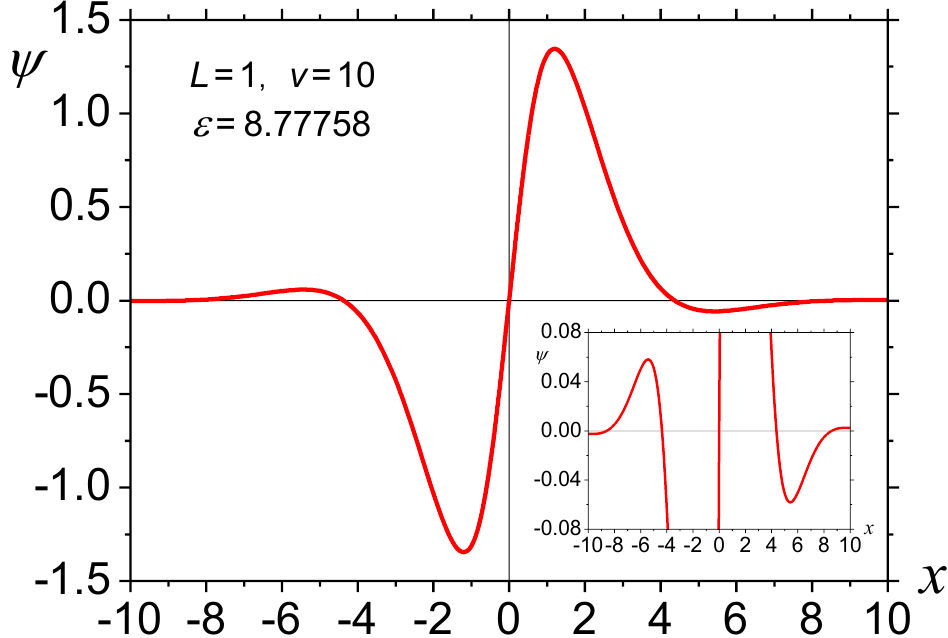}
\caption{The wave function of the ground state (left panel) and the first excited state (right panel) for the potential well.}
\label{fig:wave-functions-well}
\end{figure}

Equations (\ref{eq:positive_parity}) and (\ref{eq:negative_parity}) (or, equivalently, Eqs.(\ref{positive_parity}) and \ref{negative_parity}) are eigenvalue equations for $p L=\epsilon^{1/4}L$ defining it as a function of a dimensionless combination $v L^4$ of the depth $v$ and width $L$ of the potential well because $k L=(v L^4-(p L)^4)^{1/4}$.  The behaviour of $p L$ as a function of $\lambda=v^{1/4}L$ for states of both parities is shown in Fig.\ref{fig:pp+np-lambda}. It is not difficult to show that the number of eigenvalues $N(v,L)$ of both parities grows asymptotically
as $\simeq\lambda/\pi$ for $\lambda \to \infty$. It is clearly seen that all levels, except the ground state, start with some threshold value $\lambda_{th}$. The ground state is of
positive parity and  exists for arbitrary shallow potential well, i.e., for any $v>0$ like in the case of the one-dimensional Schrödinger's equation.

It is interesting to look at the behaviour of energy levels as a function of the width or the depth of the well when one of these parameters is fixed. Such a behaviour is shown in the middle and
right panels of Fig.\ref{fig:pp+np-lambda}. The energies of states with negative parity lie between neighbouring states of positive parity. The deeper and wider the potential well, the greater the number of eigenvalues in the well, provided that the maximal energy level does not exceed $v$, i.e., $\epsilon_{max}<v$. The ground
state exists at any finite depth or width of the potential well, whereas other levels begin at a certain threshold value of these parameters in accordance with the above-mentioned dependence of
levels on $\lambda$.

The wave functions of the ground and first excited states of the considered potential well problem are shown in the left and right panels of Fig.\ref{fig:wave-functions-well}, respectively.
Obviously, the ground state wave function is even in the coordinate $x$ (i.e., of positive parity) and the wave function of the first excited state is odd (negative parity). To demonstrate
more clearly  the presence of oscillations of wave functions in the classically forbidden region, we add inset to the left panel (ground state) and the right panel (first excited state).

Thus, the analysis of eigenfunctions of an exactly solvable square well potential problem considered in this section confirms our proposition that, for one-dimensional Hamiltonians with quartic dispersion, the node-counting rule holds only in the classically allowed region.

\section{Summary}
\label{sec:conclusions}

In the present paper, we studied  the properties of bound-state wave functions and bound-state energies of quasiparticles in weakly dispersing energy bands for 1D systems with the quartic energy-momentum dispersion and the quartic potential (the double quartic problem). To determine bound state energies, we applied the WKB semiclassical
method formulated for systems with quartic dispersion in \cite{quartic} which gives the Bohr-Sommerfeld quantization rule with non-perturbative in the Planck constant contribution.
We calculated the second and fourth WKB corrections and shown that the non-perturbative contribution, as well as  these higher
order WKB corrections are mostly relevant for the lowest bound states energies with their contributions quickly decreasing for states with higher energy.

To check the accuracy of our results found using the WKB approach, we computed the bound state energy numerically by using the variational method with the universal Gaussian basis.
Our main principal finding is that the classical oscillation theorem fails in the classically forbidden region for systems with quartic energy-momentum dispersion. We found that the bound-state
wave functions have nodes (in fact, infinitely many) in the classically forbidden region for  systems with quartic dispersion unlike the Schrödinger equation where the bound state wave functions
do not have such nodes. On the other hand, the number of nodes in the classically allowed region coincides with the quantum number $n$ of the corresponding bound state. To check this
observation,  we considered an exactly solvable potential well problem and found that the wave function of the ground state indeed does not have nodes in the classically allowed
region and oscillates producing nodes in regions beyond the potential well. The higher energy states also have nodes in the classically forbidden region whose number equals $n$.
Hence, the oscillation theorem holds in the classically allowed region in all considered examples.

The mathematical reason for the presence of nodes of the ground state wave function in the classically forbidden region is the existence of both real and imaginary parts in all four semiclassical
momenta defining the asymptotics of the wave function in the classically forbidden region. This is in contrast to the case of Schrödinger`s equation where the semiclassical momentum is purely
imaginary in the classically forbidden region which gives non-oscillating exponentially decreasing asymptotics. Oscillations in the classically forbidden region may have important physical
consequences, for example, lead to oscillating tunneling current \cite{Levitov,Sushkov}.

Since the quartic energy-momentum dispersion is obviously only the first step beyond the conventional quadratic dispersion, to fully address the properties of bound states in weakly dispersing energy bands it is important to extend the study performed here to the case of sextic and more high order dispersion. The extension to the two-dimensional case is also of interest due to the growth of the number of planar physical systems with dispersion higher than quadratic.
\vspace{3mm}

\begin{acknowledgments}
\vspace{2mm}
E.V.G. and V.P.G. acknowledge support from the National Research Foundation of Ukraine grant (2023.03/0097) “Electronic and transport properties of Dirac materials and
Josephson junctions”. B.E.G. is partially supported by the National Academy of Sciences of
Ukraine, Project No. 0122U000886.
\end{acknowledgments}

\end{document}